\documentclass[useAMS,usenatbib,usegraphicx]{mn2e}

\title[Dust properties of IRAS 18276--1431]
      {Investigation of dust properties of the proto-planetary\\
       nebula IRAS 18276--1431}
\author[K. Murakawa et al.]{
K. Murakawa$^{1}$, H. Izumiura$^{2}$, R. D. Oudmaijer$^{1}$, and
L. T. Maud$^{1}$\\ 
$^{1}$School of Physics and Astronomy, EC Stoner Building, University of Leeds,
Leeds LS2 9JT\\
$^{2}$Okayama Astrophysical Observatory, 3037-5 Honjo, Kamogata, Asakuchi,
Okayama, 719-0232 Japan\\
}
\begin{document}

\date{Accepted 1988 December 15. Received 1988 December 14; in original form 1988 October 11}

\pagerange{\pageref{firstpage}--\pageref{lastpage}} \pubyear{2002}

\maketitle

\label{firstpage}

\begin{abstract}
We investigate the circumstellar dust properties of the oxygen-rich
bipolar proto-planetary nebula IRAS 18276--1431 by means of
two-dimensional radiative transfer simulations of the circumstellar
dust shell.  The model geometry is assumed to have a torus and an
envelope which consists of a pair of bipolar lobes and a spherical AGB
shell.  The parameters of the dust and the dust shell are constrained
by comparing the spectral energy distribution (SED) and near-infrared
intensity and polarisation data with the models.  The polarisation in
the envelope reaches 50 -- 60~\% and is nearly constant in the $H$ and
$K_S$ bands in the observations.  This weak wavelength dependence of
the polarisation can be reproduced with a grain size distribution
function for the torus: 0.05~$\mu$m $\le a$ with
$n\left(a\right)\propto
a^{-\left(p=5.5\right)}\exp\left(-a/a_\mathrm{c}=0.3~\mu\mathrm{m}\right)$.
The power index $p$ is significantly steeper than that for
interstellar dust ($p\sim3$).  Similar results have also been found in
some other PPNs and suggest that mechanisms that grind down large
particles, such as sputtering, may also have acted when the dust
particles formed.  The spectral opacity index $\beta$ is found to be
0.6$\pm$0.5 from the 760~$\mu$m to 2.6~mm fluxes, which is
characterised by the dust in the torus.  This low value ($<$2)
indicates the presence of large dust grains in the torus.  We discuss
two possible dust models for the torus.  One has a size distribution
function of 1.0~$\mu$m~$\le a\le a_\mathrm{max}=5\,000.0~\mu$m with
$n\left(a\right)\propto a^{-\left(p=2.5\right)}$ and the other is
1.0~$\mu$m~$\le a\le a_\mathrm{max}=10\,000.0~\mu$m with
$n\left(a\right)\propto a^{-\left(p=3.5\right)}$.  The former has
$\beta$ of 0.633, but we are not able to find reasonable geometry
parameters to fit the SED in the infrared.  The latter has $\beta$ of
1.12, but reproduces the SED better over a wide wavelength range.
With this dust model, the geometric parameters are estimated as
follows: the inner and outer radii are 30~AU and 1000~AU and the torus
mass is 3.0~$M_{\sun}$.
Given that the torii are generally not found to be rotating, a large fraction of
the torus material is likely to be expanding.  Assuming an expansion velocity
of 15~kms$^{-1}$, the torus formation time and mass-loss rate are found to be
$\sim$300~yrs and $\sim$10$^{-2}~M_{\sun}$yr$^{-1}$ respectively.
\end{abstract}


\section{Introduction}
IRAS 18276--1431 (hereafter I18276), also known as OH 17.7--2.0, is an
oxygen-rich proto(pre)-planetary nebula (PPN).  This object has been well
studied in maser observations over the past three decades since its discovery
as a strong OH 1612~MHz source \citep{bowers78}.  No SiO maser has been detected
\citep{nyman98}.  The intensity of the H$_2$O maser dropped by a factor of
a hundred during the period between 1985 and 1990 and is now undetectable,
indicating that the mass-loss rate has rapidly decreased to below
10$^{-7}~M_{\sun}$yr$^{-1}$ \citep{engels02}.  These maser properties suggest
that the central star evolved off the asymptotic giant branch (AGB) phase very
recently.

The dust shell of this object has also been the subject of several studies.
\citet[][ hereafter SC07]{sc07} presented optical {\it HST/WFPC2} images and
infrared $K_P$, $L_P$, and $M_S$ images from the Keck II telescope.
The images show a striking bipolar appearance with an extension of $\sim$2
arcsec$\times$3 arcsec.  While the extension and the appearance look similar
in these shorter wavelength ranges, the mid-infrared images at 8.59~$\mu$m,
11.85~$\mu$m, and 12.81~$\mu$m show an elongated shape along the polar
direction with a flux peak towards the central star \citep{lagadec11}.
SC07 also performed a one-dimensional (1D) double shell model that mainly fits
the mid-infrared flux.  They found that the observed fluxes between 800~$\mu$m
and 2.6~mm are roughly proportional to $\nu^2$, where $\nu$ is the frequency,
indicating the opacity coefficient $Q_\nu\sim1$.   Hence, they concluded that
the dust grains with radii $a\ga400~\mu$m could explain the flux slope at these
wavelengths.

The dust growth in the circumstellar environment is an interesting issue.
If the ejected material radially expands at typical expansion velocities of
$\sim$15~kms$^{-1}$, the dust particles may only grow a little and retain the
properties that were set in the dust formation region.  That is to say,
particles with sizes up to a few micron should exist.  This is confirmed in
several previous papers \citep[e.g.][]{jm85}.   Bipolar objects such as I18276
are expected to have an optically thick, long-lived dust torus or disc around
the central star, which is thought to have formed because of binary
interactions \citep[e.g.][]{morris87}.  In such a thick dusty region, the dust
grows in size by mutual collisions.  Although the formation of the
disc$/$torus and the dust processes therein have been studied in various object
classes, they are still under debate and present some intriguing problems.

In the present work, we investigate the dust properties in I18276's
circumstellar dust shell (CDS) by means of two-dimensional (2D) radiative
transfer calculations.  An approximate solution was obtained by SC07's 1D SED
model.  However, since I18276 exhibits a striking bipolar lobe, we explore 2D
modeling.  Furthermore, the spectral slope of the (sub)millimeter flux is often
used to detect large grains of a few hundred micron or larger.  In our modeling,
we encounter difficulties explaining the millimeter flux excess and the fluxes
at far-infrared or shorter wavelengths simultaneously with a reasonable mass of
the material in the equatorial region.  We will revisit the spectral slope
issue.

\section{Dust shell modeling}\label{dsm}
We performed 2D radiative transfer modeling of I18276's CDS.  Our aim is
to investigate the dust properties in the envelope and the equatorial region.
An analysis of the arc-like structure and the ``search light'' feature, which
are detected in previous optical and NIR images (SC07), and the formation of
bipolarity are beyond the scope of this work.  A comprehensive discussion on
the morphology can be found in SC07.  Our modeling approach is basically
similar to our previous work \citep{murakawa10a,murakawa10b}.  We used NIR
polarimetric data obtained with VLT$/$NACO, these data are particularly useful
to estimate the dust sizes. This is complemented by  SED data collected from various sources.
The results of the HST and Keck images (SC07) and the VISIR image
\citep{lagadec11} are weighted less to constrain the parameters
in our 2D modeling.  In the following subsections, the archival data, model
assumptions, and the results for the selected model are presented.

\subsection{Observational constraints}\label{obsconst}
\begin{figure}
  \resizebox{\hsize}{!}{\includegraphics[angle=-90]{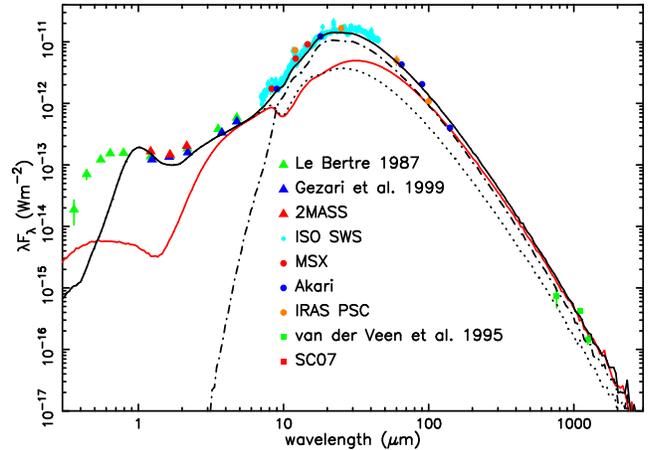}}
  \caption{The observed SED of I18276, with the model results over
           plotted.  The solid, dotted, and dashed-dotted curves
           denote the total flux, the scattered light, and the thermal
           emission, respectively.  The black lines represent the model
           with the torus and envelope.  The red line denotes the
           model with only the torus.  The sources for the photometry
           are in order of wavelength, from the optical to millimeter:
           \citet{lebertre87}, Catalog of Infrared Observations,
           Edition 5 \citep{gezari99}, 2MASS all-sky catalogue of
           point sources, Akari$/$IRC mid-IR all-sky Survey
           \citep{ishihara10}, the calibrated data (the data ID of
           10802940) of the ISO Short Wavelength Spectrometer
           \citep{kraemer02}, MSX infrared point source catalogue,
           IRAS point source catalogue, and SC07.  For the data where
           the error bars are not seen, the uncertainties are smaller than
           the plot symbols.  }
  \label{modelsed}
\end{figure}

Figure\,\ref{modelsed} shows the SED of I18276, it has a single flux peak at
$\sim$25~$\mu$m and a weak 10~$\mu$m absorption feature. In this plot,
the interstellar reddening, which is calculated to be 1.6($\pm$0.5)~mag at $V$
(SC07), is used to correct the measured fluxes for wavelengths shorter than
5~$\mu$m.

\begin{figure*}
  \centering
  \includegraphics[width=12.5cm]{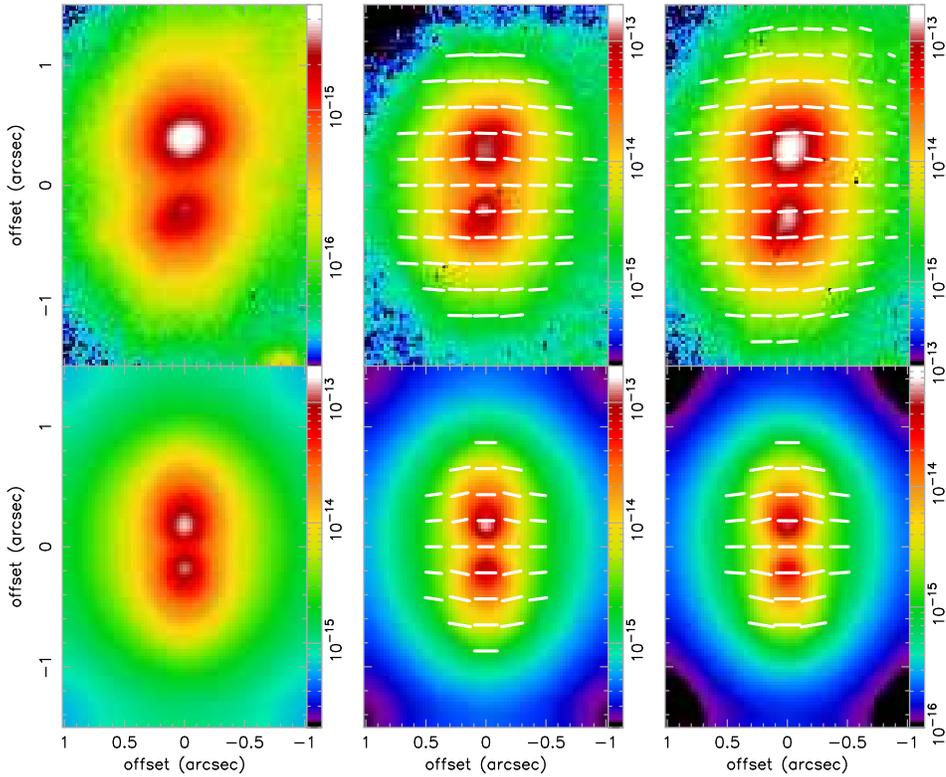}
  \caption{{\it Left three columns:} $J$, $H$, and $K_S$-band images.
           Top and bottom panels are from the VLT$/$NACO archived and our
           model results, respectively.
           The color scale bars indicate the surface brightness in
           Wm$^{-2}\mu$m$^{-1}$arcsec$^{-2}$.  In the $H$ and $K_S$-band
           images, the polarisation vector lines are overlaid.  The modelled
           images are convolved with a one-dimensional Moffat model PSF.
         }
  \label{modelimg}
\end{figure*}

The upper panels of Fig.\,\ref{modelimg} present the NIR images, which are
collected from the VLT$/$NACO data archive.  The natural seeing was 0.5 arcsec
to 0.6 arcsec in the optical during the observations, which is typical for the
Paranal site.  The $H$ and $K_S$ band data were taken with the Wollaston prism
to measure the polarisation.  The data were reduced following the same method
as before \citep{mi12}.  Standard star data were used to calibrate the
surface brightness and the polarisation, and also served to estimate the size
of the point spread function (PSF).  The measured full widths at half maximum
of the beam are 0.1 arcsec, 0.058 arcsec, and 0.063 arcsec in the $J$, $H$,
and $K_S$ bands, respectively.  The PSFs were modeled using one-dimensional
Moffat function fitting.  The PSF size 2$\sigma$ and slope $\beta$ are obtained
to be 0.146 arcsec and 1.17 in the $J$-band, 0.074 arcsec and 1.18 in the
$H$-band, and 0.048 arcsec and 1.03 in the $K_S$-band.  The model images are
convolved with the model PSF which allows comparison with the observed images.
The intensity images show a striking bipolar appearance and look similar in
the $J$, $H$, and $K_S$ bands, consistent with previous observations
(e.g.\,SC07).  For the $H$ and $K_S$ band images, the polarisation vectors are
overlaid.  The vectors are aligned perpendicularly to the bipolar axis.
The polarisation reaches 50 -- 60~\% and does not change much between the $H$
and $K_S$ bands.  There is no significant difference between the upper,
brighter and lower, fainter lobes.

In our modeling, we aim to fit the shape of the SED including the weak 10~$\mu$m
silicate feature, the constant bipolar appearance and degree of polarisation
in the NIR.

\subsection{Model assumptions}\label{modelasm}
The spectral type and the total luminosity of the central star are estimated
to be $\sim$K5 or earlier and
$\sim$$1.5\times10^4\left(D~\mathrm{kpc}/4.6\right)^2$, respectively
\citep{lebertre89}.  Previous OH maser phase-lag measurements yielded a
distance in the range 3.4 to 5.4~kpc \citep{bjs81,bjs83}.  In our model, the
central star is assumed to have a blackbody spectrum with 7000~K, which is
an intermediate value of the previous estimates of 4000~K to 10\,000~K
\citep[e.g.][]{lebertre89}.
We adopted the distance to be 3~kpc following SC07, yielding a luminosity
of 6380~$L_{\sun}$. We will start with this value, but will eventually use 8500~$L_{\sun}$ to better fit the SED.

\begin{figure}
 \centering
  \includegraphics[width=6.5cm]{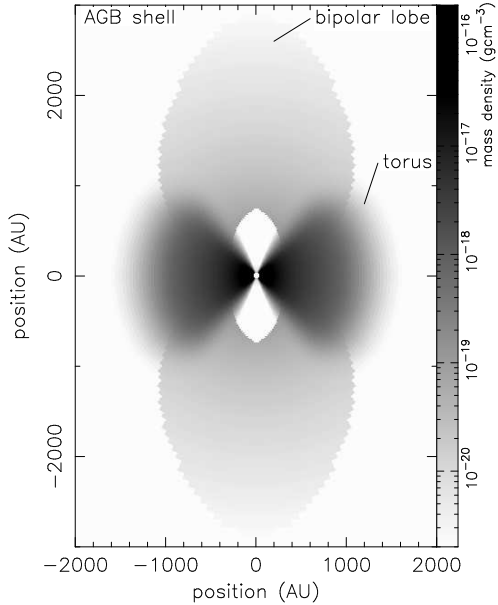}
  \caption{Mass density distribution of the model in a plane
           cutting through the symmetry axis.  The parameters are from the
           selected model (see Table~\ref{model_parm}).  The AGB shell
           is not explicitly indicated, because the mass density is lower than
           2$\times$10$^{-21}$~gcm$^{-3}$, but it exists outside the bipolar
           lobe and torus.
         }
  \label{rho}
\end{figure}

In bipolar PPNs such as I18276, it is expected that the CDS has an optically
thick dust torus$/$disc in the equatorial region and an envelope, which
characterises the lobe shape.  In our separate paper on the bipolar PPN
IRAS 16342--3814, we needed an additional optically thick, geometrically thin
disc inside the torus \citep{mi12}.  For I18276, we tried both model
geometries, but found that the inner disc is not necessary.  The mass density
form for I18276 is assumed to be
$\rho=\rho_\mathrm{torus}+\rho_\mathrm{lobe}+\rho_\mathrm{AGB}$.
We apply a mass density form $\rho_\mathrm{torus}\left(r,z\right)=
\rho_\mathrm{t}\left(r/R_\mathrm{torus}\right)^{-2}
\exp\left(-z^2/2\left(H_0r\right)^2\right)$ for the torus, where
$\left(r,z\right)$ are the two-dimensional cylindrical coordinates, and
the torus region is defined as
$R_\mathrm{in}\le \sqrt{r^2+z^2}\le R_\mathrm{torus}$.
We would like to stress that the dust particle motions in the torii of evolved
stars are not well known. For example, it is not clear whether they are
accreting onto the star, expanding, or  rotating in a Keplerian fashion, or not.
Hence, we regard the torus just as a tapered dust structure.  For the envelope,
we assume a spherical AGB shell in the outer part and a pair of bipolar lobes
inside the AGB shell.  This is because the mass-loss rate rapidly increases at
the very end of the thermal pulsing AGB phase.  The lobe is then carved out by
the fast wind  when the star evolves to the post-AGB phase.  The bipolar
appearance is mainly characterised by the lobe structure.  The mass density
forms of the lobe and AGB shell are given by equations (4) and (5) in our
I16342 paper, respectively.  They are determined by the density coefficient,
$\rho_\mathrm{e}$, the density ratios $\epsilon_\mathrm{in}$ and
$\epsilon_\mathrm{rim}$, for the inner cavity and the rim of the lobe;
the lobe shape, $\beta$, the lobe thickness, $\gamma$, the lobe size,
$R_\mathrm{lobe}$, and the outer radius $R_\mathrm{out}$.  The mass density
coefficients, $\rho_\mathrm{t}$ and $\rho_\mathrm{e}$, are determined from
the masses of the torus and envelope (i.e.\,lobe and AGB shell) assuming
a gas-to-dust mass ratio of 200.  The important parameters are the radii of
the inner $R_\mathrm{in}$ and outer $R_\mathrm{torus}$ boundaries of the torus,
the aspect ratio of the torus height $H_0$, the masses of the torus
$M_\mathrm{torus}$ and the envelope $M_\mathrm{env}$, $\gamma$,
$\epsilon_\mathrm{in}$, and $\epsilon_\mathrm{rim}$ are determined in our
modeling.  The other parameters such as $\beta$, $R_\mathrm{lobe}$, and
$R_\mathrm{out}$ are set to be fixed values.  Figure\,\ref{rho} shows the mass
density map of the inner 2000$\times$3000~AU region.

\begin{table}
  \begin{center}
  \caption[]{Model parameters of I18276's circumstellar dust shell.}
  \label{model_parm}
  \begin{tabular}{lll}
  \hline
  \hline
  parameters          & values                   & comments$^1$ \\
  \hline
  \multicolumn{3}{c}{central star} \\
  $T_\mathrm{eff}$    & 7000~K                   & \cite{lebertre89} \\
  $L_\star$           & 8500~$L_{\sun}$          & adopted$^2$  \\
  $D$                 & 3.0~kpc                  & SC07         \\
  $R_\star$           & $4.38\times10^{13}$~cm   & calculated   \\
  \hline
  \multicolumn{3}{c}{torus} \\
  $R_\mathrm{in}$     & 30~AU                    & 20 -- 30     \\
  $R_\mathrm{torus}$  & 1000~AU                  & 700 -- 1500  \\
  $H_0$               & 1.0                      & 0.8 -- 1.2   \\
  $M_\mathrm{torus}$  & 3.0~$M_{\sun}$           & 2.5 -- 4.0   \\
  $a_\mathrm{min}$    & 1.0~$\mu$m               & adopted      \\
  $a_\mathrm{max}$    & 10\,000.0~$\mu$m         & $\ga1000.0$  \\
  $\tau_\mathrm{5\mu m}$  & 52                       & calculated   \\
  \hline
  \multicolumn{3}{c}{bipolar lobe and AGB shell} \\
  $R_\mathrm{lobe}$   & 3000~AU                  & adopted$^3$  \\
  $\beta$             & 1.0                      & adopted      \\
  $\gamma$            & 0.75                     & 0.7 -- 0.8   \\
  $\epsilon_\mathrm{in}$  & 0.01                 & $<0.1$       \\
  $\epsilon_\mathrm{rim}$ & 10                   & 10 -- 15     \\
  $R_\mathrm{out}$    & 90\,000~AU               & adopted      \\
  $M_\mathrm{env}$    & 1.2~$M_{\sun}$           & 1.0 -- 1.5   \\
  $a_\mathrm{c}$      & 0.3~$\mu$m               & 0.3 -- 0.4   \\
  \hline
  \end{tabular}
  \end{center}
  $^1$ Ranges give the uncertainty of the corresponding model parameters.
  $^2$ Based on comparison of the SED.
  $^3$ Based on comparison with the intensity image.
\end{table}

The mineralogy of dust in the CDS is not well known.  We therefore
simplify the dust model and adopt the optical constants of oxygen
deficient silicate \citep{ossenkopf92}.  We assume different dust
sizes for the torus and the envelope.  The dust size in the envelope
can be constrained with polarimetric data.  We assume a KMH-like size
distribution function \citep{kmh94}: $n\left(a\right)\propto
a^{-p}\exp\left(-a/a_\mathrm{c}\right)$ with $a_\mathrm{min}$ of
0.05~$\mu$m and note that a steep power index $p$ of 5.5 results in a
weak wavelength dependence.  Only {\bf the cutoff size, $a_\mathrm{c}$,} is a free parameter.
The wavelength dependency of the power index $p$ will be discussed in
Sect.\,\ref{dustlobe}.

The dust sizes in the torus can be constrained with the spectral flux slope
$\alpha$ in the (sub)millimeter wavelength ranges. With the flux data between
761~$\mu$m and 2.6~mm, $\alpha$ is 2.6$\pm$0.5, which yields the spectral
opacity index $\beta$ of 0.6$\pm$0.5 ($=\alpha-2$).  In principle, this allows
us to determine a size distribution function, however, it is not a unique
solution.  We encounter difficulties in fitting the SED at the shorter
wavelengths with the derived dust models.  The details of this problem will be
discussed in sect.\,\ref{dusttorus}.  Instead of applying a dust model that can
be determined using $\beta$, we attempted to find one that fits the SED better
over a wider wavelength range.  Assuming an MRN-like size distribution function
\citep{mrn77}:
$n\left(a\right)\propto a^{-3.5}$ with $a_\mathrm{min} \le a\le a_\mathrm{max}$
, $a_\mathrm{max}$ is determined by the SED fitting.  For $a_\mathrm{min}$,
we adopt 1.0~$\mu$m.   However, since we do not have enough clues to determine
the value precisely, we exclude the $a_\mathrm{min}$ value from the discussion
of the dust properties in the torus in Sect.\,\ref{dusttorus}.

\subsection{Results and comparison with the observations}\label{model}
The aforementioned parameters are estimated by means of radiative transfer
calculations using our \textsf{STSH} code \citep{murakawa08a}.  We first
tried several parameter sets to get approximate solutions and to understand
the parameter dependence on the model results.  Then, a few thousand parameter
sets were examined by SED fitting.  From this result, a dozen good parameter
sets are chosen and their images are evaluated.  We finally chose one that
explains the observations fairly well.  We would like to stress that it is
very difficult to find a parameter set that reproduces all aspects of the
observations.  Because our aim is to investigate the dust properties, our
modeling is mostly concerned with explaining the NIR polarisation and the SED
over a wide wavelength range.  The images in the optical and MIR are not
accurately reproduced.  The derived parameters are:
$R_\mathrm{in}=30$~AU, $R_\mathrm{torus}=1000$~AU,
$M_\mathrm{torus}=3.0~M_{\sun}$, $a_\mathrm{min}=1.0~\mu$m,
$a_\mathrm{max}=10\,000.0~\mu$m, the optical depth in the torus midplane is
52 at 5.0~$\mu$m, $M_\mathrm{env}=1.2~M_{\sun}$, and
$a_\mathrm{c}=0.3~\mu$m.  The viewing angle measured from the polar axis
is $80\degr$.  Table \ref{model_parm} lists these parameters and uncertainties.
We briefly summarise the selected model results in this section and discuss
the effects of dust sizes on the model results in detail in
sect.\,\ref{dustprop}.

The SED of the selected model is plotted with black lines in
Fig.\,\ref{modelsed}.  The aperture sizes used for the photometric and
spectroscopic data are different for the various observations; they are
4 arcsec in radius (for 2MASS) or larger.  Because the apparent size of the
nebulosity is small and most flux is concentrated within $\sim$1 arcsec,
the effect of aperture
size on the observed SED is negligible in I18276.  In fact, the model flux
derived applying a 4 arcsec aperture differs less than 1~\% from that applying
an infinite aperture from optical to millimetre wavelengths.  The model fits
the observed SED well apart from the optical.  The weak 10~$\mu$m feature is
due to a filling in the underlying absorption with strong thermal emission.
In fact, the MIR images show a flux peak towards the central star
\citep{lagadec11} and the calculated optical depth becomes high as
41 at 10~$\mu$m.  The red line denotes the result of the selected model but
with only the torus, e.g.\,$M_\mathrm{env}$ is simply set to be 0.
It approximately indicates the contribution from the torus in the selected
model.  In the wavelengths shorter than $\sim$9~$\mu$m and between
$\sim$9~$\mu$m and $\sim$200~$\mu$m, the scattered light and thermal emission
from the envelope, respectively, dominate.  The SED of the selected model fits
the observations comparably well in the NIR and longer wavelengths.  However, a large
discrepancy is seen in the optical.  In some models with other parameter sets,
e.g.\,$a_\mathrm{min}=5~\mu$m dust for the torus and a shallower power index of
the mass density distribution in the envelope, we find that these models can
fit the SED in the optical better than the selected model.  However, these
models provide poorer fits to the images or  the MIR fluxes.

The bottom panels in Fig.\,\ref{modelimg} show the model results.  The pair of bipolar
lobes and a nearly constant appearance from $J$ to $K_S$ bands are reproduced.
The optical and MIR images are also reproduced although they are not presented.
The bipolarity is seen in the optical, $L_P$, and $M_S$ bands.
On the other hand, the 11.85~$\mu$m image shows a single peak.  These results
are qualitatively consistent with the observations
\citep[SC07 and][]{lagadec11}.  However the details of the surface brightness
structures of the images are not accurately reproduced because the images at
these wavelengths are not used to strictly constrain the model parameters.

The polarisation vectors are aligned perpendicular to the bipolar axis in the
entire nebulosity.  Since a centro-symmetric pattern is seen in the raw images,
this alignment is due to the PSF smoothing effect, as is also demonstrated in
our I16342 paper \citep{mi12}.  The polarisation in the bipolar lobes reaches
50 -- 65~\%, which is slightly higher than the observations.
If $a_\mathrm{c}=0.4~\mu$m is used, the polarisation becomes 45 -- 60~\%.
We confirm that our estimation of the particle size in the envelope is
consistent with that by \citet{gledhill05}.

We reach an important conclusion with respect to the particle sizes.
{\bf
In the envelope, the mass fraction of the particles with radius larger than
the cutoff size ($a_\mathrm{c}=0.3~\mu$m) is 28~\%.  This fraction decreases dramatically
for larger sizes, e.g.\,only 1.7~\% for $a\ge1.0~\mu$m.
The submicron sized dust dominates in the envelope.  On the other hand,
the fraction of the particles larger than 400~$\mu$m is high as 81~\% for
the torus.  Hence, the particles in the torus are expected to be significantly
larger than in the envelope.
}
In Sect.\,\ref{dustlobe}, we discuss
the effect of the power index, $p$, on the polarization and provide our
interpretation on the derived $p$ of 5.5.  In Sect.\,\ref{dusttorus},
we discuss the determination of the particle sizes using the millimeter flux
excess and the problems encountered in our modeling.

\section{Dust properties}\label{dustprop}
\subsection{In the bipolar lobe and the AGB shell}\label{dustlobe}
In NIR polarimetric images of bipolar reflection nebulae such as those
seen around for example evolved stars, young stellar objects (YSOs),
and active galactic nuclei, the polarisation vectors are
centro-symmetrically aligned and the degree of linear polarisation
reaches $\ga10$~\% in general.  This result is explained by single
scattering off submicron-sized dust particles.  The polarisation is
sensitive to the dust size, or to be more precise, the size parameter
$x=2\pi a/\lambda$ for $x\sim1$.  In the case of YSOs, a strong
wavelength dependence of the polarisation is seen which can be explained
with a size distribution function: $n\left(a\right)\propto a^{-3.5}$
\citep[e.g.][]{hales06,beckford08,murakawa08a}.  This value for the
exponent can be readily understood because the dust in the YSO
envelope originates from interstellar matter.  In contrast, weak
wavelength dependencies are found in some evolved stars.  The
polarisation ranges from 35 to 60~\% in the optical to NIR in 
Frosty Leo \citep{dougados90,ss94}.  \cite{ss94} found a steeper power
index $p$ ranging from 5.2 to 5.5.

\begin{figure}
  \resizebox{\hsize}{!}{\includegraphics[angle=-90]{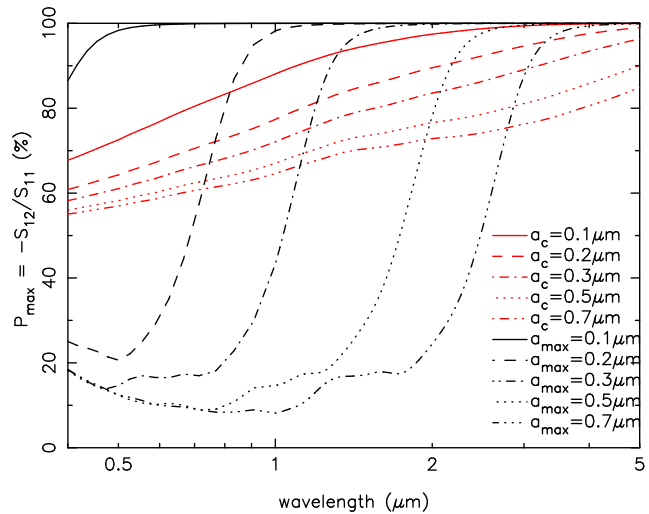}}
  \caption{Plot of scattering property.  The degree of polarisation is
           the maximum values of the single scattered light.  Red and black
           lines denote the results of G1 and G2, respectively.
         }
  \label{scat}
\end{figure}

We now investigate the effects of the dust sizes and the size
distribution function on polarisation.  Two models are considered: G1
has 0.05~$\mu$m $\le a$ and $n\left(a\right)\propto
a^{-5.5}\exp\left(-a/a_\mathrm{c}\right)$, and G2 has 0.005~$\mu$m
$\le a\le a_\mathrm{max}~\mu$m and $n\left(a\right)\propto a^{-3.5}$.
Upper limits of the dust size $a_\mathrm{c}$ or $a_\mathrm{max}$ of
0.1, 0.2, 0.3, 0.5, and 0.7~$\mu$m are examined.  Figure~\ref{scat}
presents the resulting polarisation.  The degree of polarisation
$P_\mathrm{max}$ is calculated  using the ratio of $-S_{12}$ to $S_{11}$ at
the scattering angle that maximises the polarisation, where $S_{11}$
and $S_{12}$ are the scattering matrix elements for spherical grains
\citep{bh83}.  These values should be the upper limit of the
polarisation expected in the envelope because the scattered light at
various scattering angles is integrated in the observed data.  In
fact, the polarisation values in our model are 50 -- 65~\% in the $H$
and $K_S$ bands, but the above calculations give 70 -- 80~\%.

G2 has a strong wavelength dependence in the optical to the NIR, as is
observed in YSO envelopes.  On the other hand, G1 has a weak
dependence in wavelength and dust size.  The scattering properties are
determined by the relative fractions of small and large dust.  The
scattering matrix elements $S$ are weighted with $n\left(a\right)$ by
$<S>=\int Sn\left(a\right)da/\int n\left(a\right)da$.  In G2 the
scattering property is dominated by the dust at the maximum sizes.
The polarisation varies rapidly around $x\sim1$,
e.g.\,$\lambda\sim1~\mu$m for $a_\mathrm{max}=0.3~\mu$m.  It is of
interest that in fact the polarisations cannot be simultaneously fit
in the $H$ and $K_S$ bands if the G2 dust is applied in the envelope
in our modeling.  For example, whereas the $a_\mathrm{max}=0.4~\mu$m
model reproduces the $H$ band polarization with a value of
$P_\mathrm{H}\sim55$~\%, the model returns a too high $K$ band
polarization of $\sim70$~\%. The 0.5~$\mu$m dust models, instead,
reproduce the $K$ band polarization and return
$P_\mathrm{K}\sim60$~\%, however they result in too low
$P_\mathrm{H}\sim27$~\%.  We see large discrepancies despite a
small difference of $\sim$0.1~$\mu$m in the $a_\mathrm{max}$ values.
On the other hand, the G1 model has a larger fraction of small
particles ($x<1$).  Thus, the polarisations are higher even at shorter
wavelengths, i.e.\,$\lambda<1~\mu$m.  In contrast, lower polarisations
are obtained at longer wavelengths ($\lambda>1~\mu$m) because of the
large particles, even though the fractions are low.

It is also useful to note that the polarisations of the G1 models are lower
than those of G2 at $\lambda>3~\mu$m even with the same upper limit sizes.
This is because dust models with the KMH-like size distribution function
have particles with sizes exceeding $a_\mathrm{c}$.
If a steeper power index, e.g.\,$p=5.5>3.5$, is assumed in the MRN-type dust
model, the scattering properties at longer wavelengths
($\lambda\ga2\pi a_\mathrm{max}$) consequently becomes similar to those with
a shallower power index.  We reach two conclusions.  First, in objects that
show a weak wavelength dependence in their envelopes such as I18276 and Frosty
Leo, a steep power index $p(>3.5)$ is expected in the size distribution
function.  Second, if the polarisation is measured over a wider wavelength
range such as from the optical to 3 to 5~$\mu$m, which is still in the scattering
regime in cool dust regions, the size distribution functions are better
constrained, e.g.\,the lower limit and whether the MRN-type or the KMH-type
is more favoured.

This brings us to the more general question whether the dust
properties depend on the objects under consideration. A NIR
polarimetric survey of a dozen of PPNs was presented
\citep{gledhill01}.  Four objects (IRAS 17436+5003, IRAS 19114+0002,
IRAS 18095+2704, and IRAS 19500--1709) out of seven for which the
nebulosity is spatially resolved, show nearly constant polarisations
in the $J$ and $K$ bands.  Including three other objects (Frosty Leo,
I16342, and I18276), a weak wavelength dependence on polarisation is
detected.  We thus find similar polarisation behaviour for different
morphologies, i.e.\,spherical, elliptic, and bipolar, while all
objects except for IRAS 19500--1709 are oxygen-rich.  Differences in
morphology are thought to result from the different initial stellar
mass, i.e.\,the higher the mass, the more asymmetry
\citep[e.g.][]{cs95}.  These observational results suggest that the
initial stellar mass and the dust shell morphology do not play
important roles in determining the dust size distribution function.
The power index of $\sim$3 that is found in the interstellar medium
\citep{mrn77} is thought to be the result of dust formation process of
mass-losing evolved stars \citep[e.g.][]{bh80}.

\cite{dgs89} analysed the dust size spectrum in the dust-driven winds of carbon
rich stars.  In their models, dust particles form via nucleation at
$\la2.0~R_\star$, the size distribution function rapidly approaches the final
distribution within $\sim$20~$R_\star$.  They found that the size distribution
functions are proportional to $a^{-5}$ in the case without drift between the
gas and dust motions and to $a^{-7}$ in the case with drift.  Their results show
that large drift velocities steepen the size distribution function because they
can break up large particles by sputtering.  \citet{woitke06} and \citet{ha07}
pointed out that in oxygen-rich winds, the radiation pressure on the dust is too
weak to drive the wind because of low dust opacities.  In fact, survey
observations of a large number of AGB stars and mass-loss models have shown
that expansion velocities are lower in oxygen-rich stars
\citep{groenewegen99,so01,olofsson02}, suggesting lower drift velocities in
oxygen-rich stars than the carbon-rich objects.  At first glance, this is in
the opposite sense to the result that the majority of the objects where steeper
power indices are found are oxygen-rich in the aforementioned polarisation
survey.  Follow up observations of various objects, i.e.\,C-$/$O-rich, and
various stellar luminosities, mass-loss rates, and evolutionary stages, and
further analyses of the dust formation and stellar wind are essential.
Although we are not in a position to identify the physical reasons, one may
expect that shattering or sputtering works efficiently in the circumstellar
environments in objects like I18276.

\begin{figure}
  \resizebox{\hsize}{!}{\includegraphics{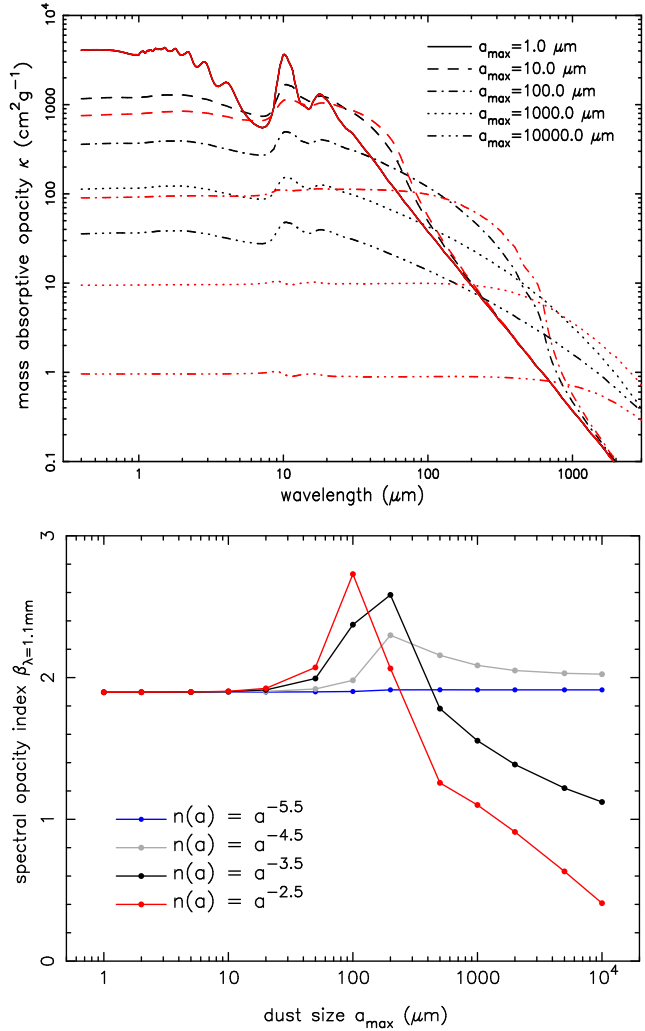}}
  \caption{{\it (top)} mass absorptive opacity $\kappa\left(\lambda\right)$
           and {\it (bottom)} spectral opacity indices $\beta$ at
           $\lambda=1.1$~mm as function of the maximum size $a_\mathrm{max}$.
           Black and red lines denote dust models with a power index $p$ of
           3.5 and 2.5, respectively.  The minimum size $a_\mathrm{min}$ is set
           to be 1.0~$\mu$m.  In the bottom panel, the dust models with a power
           index of 4.5 (gray) and 5.5 (blue) are also indicated.  In the top
           panel, the plot for the $a_\mathrm{max}=1.0~\mu$m model is identical
           to the single sized dust with a radius of 1.0~$\mu$m.
         }
  \label{kappa}
\end{figure}

\begin{table}
  \begin{center}
  \caption[]{Dust model parameters for the torus}
  \label{dust_parm}
  \begin{tabular}{lllllll}
    \hline
    ID & $a_\mathrm{max}/p$ & $\beta$  & $\kappa$ & $C_\mathrm{ext}$
       & \multicolumn{2}{c}{$M_\mathrm{torus}$} \\
    \hline
       & $\mu$m$/$ & & cm$^2$g$^{-1}$ & cm$^2$g$^{-1}$
       & \multicolumn{2}{c}{$M_{\sun}$} \\
    \hline
    A  & 10\,000.0$/$3.5 & 1.12  & 1.44  &   103  &   3.0  & 2.11 \\
    B  &    5000.0$/$2.5 & 0.633 & 1.18  &   4.54 &  68.1  & 2.59 \\
    \hline
  \end{tabular}
  \end{center}
  The spectral opacity index $\beta$ and the mass absorptive opacity $\kappa$
  are values at $\lambda=1.1$~mm.  The value of the extinction cross section
  $C_\mathrm{ext}$ is for $\lambda=5.0~\mu$m.  For the torus mass
  $M_\mathrm{torus}$, the values are obtained to keep a constant optical
  depth at 5.0~$\mu$m $\tau=52$ (left column),
  i.e.\,$C_\mathrm{ext}M_\mathrm{torus}$ is constant, and to fit the flux
  at $\lambda=1.1$~mm (right column).
\end{table}

\subsection{In the torus}\label{dusttorus}
Measuring the flux excess at submillimeter to millimeter wavelengths
provides a simple and powerful method to evaluate whether large dust grains
($\ga$ a few hundred microns) exist in the disc$/$torus region or not
\citep{bs91,mn93,draine06}.  In fact, numerous previous works have found
dust growth particularly in discs around young stars
\citep[e.g.][]{calvet02,testi03,lommen07}. Some intensive modeling
to investigate the effects of the dust growth on the SEDs and images has been
done \citep[e.g.][]{dalessio01}.  In our modeling of I18276, the flux from the
torus dominates in the FIR or longer wavelengths (see Fig.\,\ref{modelsed} and
Sect.\,\ref{model}).  In this result, the dust size in the torus was estimated
to fit the SED in a wide wavelength range.  In this section, we discuss dust
models that fit the flux slope in the submillimeter and millimeter wavelength
ranges by applying the aforementioned analysis.

The spectral opacity index cannot be directly determined from observations.
Rigorously speaking, it should be derived from the (sub)millimeter flux
by solving the radiative transfer problem.  In special cases, such as accretion
discs \citep{bs91}, approximate solutions are given.  Here we show a simpler
prescription.  Let us assume that the flux in that wavelength regime is
dominated by  emission from the torus and that this emission can be
approximated  by blackbody radiation with a representative dust temperature,
$B_\nu\left(T_\mathrm{d}\right)$, the flux $F_\nu$ is given by
$F_\nu\approx M_\mathrm{torus}\kappa_\nu B_\nu\left(T_\mathrm{d}\right)
D^{-2}\zeta^{-1}$, where $\kappa_\nu$, $D$, and $\zeta$ are the mass absorptive
opacity, the distance to the object and the gas-to-dust mass ratio,
respectively.  If the dust emission at this wavelength range is assumed to be
in the Rayleigh-Jeans limit, the blackbody function can be approximated by 
a power law, $B_\nu\left(T_\mathrm{d}\right)\propto\nu^2$.  In addition,
if the dust opacity has a power-law dependence on frequency,
$\kappa_\nu\propto\nu^\beta$, the flux is given by $F_\nu\propto\nu^\alpha$
with $\alpha\equiv d\ln F_\nu/d\ln\nu =\beta+2$.  This can be obtained when
the ratio of the optically thick to thin emission, $\Delta$, is neglected in
the result derived by \citet{bs91}.

With the observed flux data of I18276 from $\lambda=761~\mu$m to 2.6~mm,
$\beta$ is found to be 0.6$\pm$0.5 where the flux measurement errors are taken
into account.  Next, we consider $\beta$ in the dust models.
Figure\,\ref{kappa} shows the mass absorptive opacity $\kappa$ and the spectral
opacity index $\beta$ at $\lambda=1.1$~mm for several spherical dust models,
which are calculated using Mie scattering theory.  The examined size
distribution functions have an MRN-like form with $p=5.5$, 4.5, 3.5 (G3),
and 2.5 {\bf (G4)}. The minimum size $a_\mathrm{min}$ is set to be 1.0~$\mu$m and
the maximum sizes that were examined range from 1.0~$\mu$m to 10\,000~$\mu$m.
$\beta$ is constant at $\sim2$ for $a_\mathrm{max}\le100.0~\mu$m, as is known
for IS dust.  The behaviour of $\beta$ does not change much for
$a_\mathrm{min}\la10~\mu$m.  An anomalous $\beta>2$ is seen at 100~$\mu$m
$\le a_\mathrm{max}\le400~\mu$m.  This is due to a sudden drop in the opacity
at $x\sim1$, as seen in the top panel.  For $a_\mathrm{max}\ga400~\mu$m,
$\beta$ monotonically decreases.  The behaviour of $\beta$ depends on the power
index $p$. The absorptive opacity is $\pi a^2Q\left(a\right)$, where
$Q\left(a\right)$ is the absorptive coefficient.  The fraction of dust opacity
to grains with size $a$ or larger, $f\left(a\right)$, is given by
$f\left(a\right)\ga\left(a^{3-p}_\mathrm{max}-a^{3-p}\right)/
\left(a^{3-p}_\mathrm{max}-a_\mathrm{min}^{3-p}\right)$.  This should be the
lower limit because $Q$ depends on $a$ and is less than unity at smaller $a$.
The larger the $p$ value, the smaller the fraction of large particles.  In the
dust models with $p=5.5$ and 4.5, $\beta$ stays $\sim$2 even with large
$a_\mathrm{max}$ values.  In the G3 type, $\beta$ decreases down to $\sim1$.
Such a lower limit is reproduced in astronomical silicate particles
\citep[$\beta_\mathrm{min}\sim0.85$,][]{draine06}.  A very low $\beta<1$ in
an oxygen-rich environment implies a lower $p<3.5$, as seen in G4.  We find
that a dust model with $a_\mathrm{max}=5000~\mu$m and $p=2.5$ has
$\beta=0.633$, which agrees with the observations.  We assume the
representative temperature of 200~K, which actually varies between 100~K and
700~K in the torus midplane, a distance of 3~kpc, and a gas-to-dust mass ratio
of 200.  To fit the (sub)millimeter flux, the torus mass should be
2.59~$M_{\sun}$ for the observed flux of 156~mJy at $\lambda=1.1$~mm
\citep{vanderveen95} and the dust opacity at this wavelength of
1.18~cm$^{-2}$g$^{-1}$.

With this dust model, we encounter the serious problem that the NIR flux
cannot be reasonably fit.  The reason is that the extinction cross section of
this dust model is too low in the NIR.  It would be easier to understand this
using our selected model as reference.  The optical depth of the torus is found
to be 52 at 5.0~$\mu$m.  To fit the NIR flux with a dust model other than that
used for the selected model, the optical depth of the torus should be kept
constant.  Namely, $C_\mathrm{ext}M_\mathrm{torus}$ should be constant because
$\tau_\mathrm{torus}\propto C_\mathrm{ext}M_\mathrm{torus}$, where
$C_\mathrm{ext}$ is the extinction cross section at $\lambda=5~\mu$m.
Table\,\ref{dust_parm} lists $\beta$, $\kappa$, $C_\mathrm{ext}$, and torus
masses of two dust models.  Whereas $C_\mathrm{ext}$ of the dust model A
(the selected model shown in Figure\,\ref{modelsed}), is high at
103~cm$^2$g$^{-1}$, that for model B, which fits $\beta$, is only
4.54~cm$^2$g$^{-1}$.  The torus mass for model B is required to be
unreliably large at 68.1~$M_{\sun}$.  Hence, the $\beta$ value is more
likely close to unity.

In either estimations ($M_\mathrm{torus}=3.0~M_{\sun}$ or 2.59~$M_{\sun}$),
the torus is massive.  Even if a binary companion with mass of
$\sim$1~$M_{\sun}$ is assumed to exist in I18276's stellar system,
it is unlikely that the derived torus has a pure Keplerian rotating motion.
Rather a large fraction of the torus material should be radially expanding.
A hydrodynamic simulation of a binary interacted outflow model shows a spiral
structure in AGB shells even in the models that are classified as bipolar
\citep{mm99}.  Similar features are detected in some AGB shells, e.g. NGC~3068
\citep{mh06} and R~Scl \citep{maercker12}, although these are optically thin
cases.  In addition, up to now, a Keplerian rotating motion is detected only
for the Red Rectangle \citep{bujarrabal03} while other objects are found to
have expanding motions for which the velocity is well studied from
observations.  Assuming an expansion velocity, $V_\mathrm{exp}$, of
15~kms$^{-1}$, which is an intermediate value of 12 -- 17~kms$^{-1}$
\citep[][ SC07]{bjs83,heske90}, the torus is formed in $\sim$300 years
($=R_\mathrm{torus}/V_\mathrm{exp}$).  To accumulate a mass of 3.0~$M_{\sun}$,
the mass-loss rate should be high as $\sim$10$^{-2}~M_{\sun}$yr$^{-1}$.
Large mass-loss rates lasting a short time at the very end of the AGB phase
have been found in some other objects \citep[e.g.][]{bujarrabal01,sc04}.

We briefly highlight the problem of dust growth in the torus.  In some objects,
for which large grains are detected, e.g.\, the Red Rectangle
\citep{jura95,vanwinckel98} and AFGL~2688 \citep{bn96,jura00}, a long-lived
disc is also expected to be present.  This is a quite natural expectation
because the dust particles have a larger chance to collide and grow in size
in a disc than in an expanding outflow, although this is still hypothesised.
However, as mentioned above, only for the Red Rectangle a stable rotating disc
is detected.  The Red Rectangle's Keplerian rotating disc is spatially resolved
by CO emission line radio interferometry \citep{bujarrabal05}.  These authors
estimate a disc mass of 0.04~$M_{\sun}$, a radius of 560~AU, and a central mass
of $\sim$0.9~$M_{\sun}$.  This system is somewhat similar to the solar nebula
besides the stellar luminosity.  Hence, the dust growth in Red Rectangle can be
explained with  theories developed for the solar-nebula
\citep[e.g.][]{weidenschilling80,nakagawa86}.  On the other hand, in some
objects like I18276 (this work) and IRAS 22036+5306 \citep{sahai06}, in which
large fractions of large grains are expected in the torus, the torus mass is
very high ($\ga1~M_{\sun}$) and expansion motions are detected.  Further
studies, in which the dust growth theory should be optimised for the
circumstellar environment around evolved stars, are encouraged to better
understand such massive large particles in the tori of these objects.

\section{Conclusion}
We modeled the CDS of the oxygen-rich bipolar PPN IRAS 18276--1431 by means of
two-dimensional dust radiative transfer calculations.  The previously observed
SED and NIR polarimetric images were used to constrain the physical
parameters.  The primary aim of our work was to investigate the dust properties.

The polarisation reaches 50 -- 60~\% in the bipolar lobes and is almost constant
in the $H$ and $K_S$ bands.  This weak wavelength dependence and high values
can be reproduced by dust models with a steeper power index of 5.5 of the size
distribution function than interstellar dust.  If the dust in I18276's
envelope has conserved its properties since its formation during the AGB phase,
mechanisms that break up large grains and increase the fraction of small
particles, such as sputtering, may  have also acted.  Polarization measurements
at a large wavelength range for various object classes of C-rich and O-rich,
and AGB stars to PNs will provide better constraints of the size distribution
functions and will help our understanding of the dust formation and process
in circumstellar environments.

We also investigated the dust properties in the torus.  Two candidate dust
models were examined.  One has a size distribution function of
$n\left(a\right)\propto a^{-2.5}$ and $a_\mathrm{max}=5\,000~\mu$m and
the spectral opacity index $\beta$ of 0.633.  This model agrees with the
observed $\beta$ of 0.6 from the 760~$\mu$m to 2.6~mm fluxes, however,
it reproduces too low infrared flux because the opacity is too small.
The other dust model has $n\left(a\right)\propto a^{-3.5}$ and
$a_\mathrm{max}=10\,000~\mu$m and fits the SED over a wide wavelength range
besides the (sub)millimeter opacity slope ($\beta=1.12$).
The torus masses are large for both models;  2.59~$M_{\sun}$ (former)
and 3.0~$M_{\sun}$ (latter).  In this massive torus, the material is probably
radially expanding rather than purely rotating around the central star, as
is expected in the previous CO emission line observations.  If the expansion
velocity of 15~kms$^{-1}$, which is an intermediate value of the previous
estimations, is assumed, the torus has been formed in $\sim$300~yrs with
a mass loss rate of $\sim$$10^{-2}~M_{\sun}$yr$^{-1}$.

\section*{Acknowledgments}
The NIR polarimetric images are obtained at the European Southern Observatory
(proposal ID: 075.D-0268).  The photometric and spectroscopic data are data
produces from the Two Micron All Sky Survey, which is a joint project of the
University of Massachusetts and the Infrared Processing and Analysis
Center$/$California Institute of Technology, funded by the National Aeronautics
and Space Administration and the National Science Foundation, AKARI, a JAXA
project with the participation of ESA, Infrared Space Observatory, and
Infrared Astronomical Satellite.

\end{document}